\documentclass[11pt,a4paper]{article}

\usepackage{jcappub}
\usepackage{graphicx}
\usepackage{amssymb}
\usepackage{amsmath}
\usepackage{bm}
\usepackage{latexsym}
\usepackage{epsfig}
\usepackage{psfrag}
\usepackage{ulem}
\usepackage{textcomp}
\usepackage{color}
\usepackage{pstricks}
\usepackage[utf8]{inputenc}

\makeatletter
\gdef\@fpheader{}
\makeatother

\def\nn{\nonumber} 

\def\f{\frac}
\def\l{\left}
\def\r{\right}
\def\d{{\rm d}}
\def\Mpl{M_{_{\rm Pl}}}
\def\beq{\begin{equation}}
\def\eeq{\end{equation}} 
\def\beqa{\begin{eqnarray}}
\def\eeqa{\end{eqnarray}}

\def\cN{\mathcal N}

\def\vk{{\bm k}}
\def\kT{k_{_{\rm T}}}
\def\vka{{\bm k}_{1}}
\def\vkb{{\bm k}_{2}}
\def\vkc{{\bm k}_{3}}
\def\ska{{k_{1}}}
\def\skb{{k_{2}}}
\def\skc{{k_{3}}}
\def\Mp{M_{_{\rm Pl}}}
\def\cG{{\cal G}}
\def\ei{\eta_{\rm i}}

\def\ee{\eta_{\rm e}}
\def\hnl{h_{_{\rm NL}}}

\def\nt{n_{_{\rm T}}}
\def\pt{{\mathcal P}_{_{\rm T}}}

\def\cB{{\mathcal B}}

\def\gB{\gamma^{_{\rm B}}}
\def\gBe{\gamma^{_{^{\rm B}}}}

\newcommand{\g}{\gamma}

\newcommand{\viz}{\textit{viz.~}}
\newcommand{\ie}{\textit{i.e.~}}

\begin{document}
\title{The tensor bi-spectrum in a matter bounce}
\author{Debika~Chowdhury,}
\affiliation{Department of Physics, Indian Institute of 
Technology Madras, Chennai~600036, India}
\emailAdd{debika@physics.iitm.ac.in}
\author{V.~Sreenath\footnote{Current address:~Department of Physics 
and Astronomy, Louisiana State University, Baton Rouge, LA~70803,
U.~S.~A.. E-mail:~sreenath@lsu.edu.}}
\author{and L.~Sriramkumar}
\emailAdd{sriram@physics.iitm.ac.in}
\date{today} 
\abstract{Matter bounces are bouncing scenarios wherein the universe contracts 
as in a matter dominated phase at early times.
Such scenarios are known to lead to a scale invariant spectrum of tensor 
perturbations, just as de Sitter inflation does.
In this work, we examine if the tensor bi-spectrum can discriminate between the 
inflationary and the bouncing scenarios.
Using the Maldacena formalism, we analytically evaluate the tensor bi-spectrum 
in a matter bounce for an arbitrary triangular configuration of the wavevectors.
We show that, over scales of cosmological interest, the non-Gaussianity parameter 
$\hnl$ that characterizes the amplitude of the tensor bi-spectrum is quite small 
when compared to the corresponding values in de Sitter inflation.
During inflation, the amplitude of the tensor perturbations freeze on super-Hubble
scales, a behavior that results in the so-called consistency condition relating the 
tensor bi-spectrum and the power spectrum in the squeezed limit.
In contrast, in the bouncing scenarios, the amplitude of the tensor perturbations 
grow strongly as one approaches the bounce, which suggests that the consistency 
condition will not be valid in such situations.
We explicitly show that the consistency relation is indeed violated in the
matter bounce.
We discuss the implications of the results.}
\maketitle


\section{Introduction}\label{sec:introduction}

Bouncing models correspond to situations wherein the universe initially 
goes through a period of contraction until the scale factor reaches a 
certain minimum value before transiting to the expanding phase. 
They offer an alternative to inflation to overcome the horizon problem, as 
they permit well motivated, Minkowski-like initial conditions to be imposed 
on the perturbations at early times during the contracting phase (see, for
instance, Refs.~\cite{Starobinsky:1979ty,Wands:1998yp,Finelli:2001sr,
Peter:2002cn,Peter:2003rg,Martin:2003sf,Martin:2003bp,Allen:2004vz,Martin:2004pm,
Creminelli:2004jg,Creminelli:2007aq,Cai:2007qw,Abramo:2007mp,Finelli:2007tr,
Falciano:2008gt,Cai:2011tc,Qiu:2011cy,Levy:2015awa};
for reviews, see Refs.~\cite{Novello:2008ra,Brandenberger:2012,Battefeld:2014uga}).
Interestingly, certain bouncing scenarios can lead to nearly scale invariant
perturbation spectra (see, for example, Refs.~\cite{Finelli:2001sr,Creminelli:2007aq,
Levy:2015awa}), as is required by observations~\cite{Ade:2015xua,Ade:2015lrj}.
For instance, a bouncing model wherein the universe goes through a 
contracting phase as driven by matter---referred to as a matter bounce---is 
known to lead to an exactly scale invariant spectrum of tensor perturbations 
as in de Sitter inflation~\cite{Starobinsky:1979ty,Wands:1998yp,Finelli:2001sr}.
Clearly, it will be worthwhile to examine if non-Gaussianities can help us
discriminate between such scenarios~\cite{Cai:2009fn,Gao:2014hea,Gao:2014eaa}.

\par

The most dominant of the non-Gaussian signatures are the non-vanishing 
three-point functions involving the scalars as well as the 
tensors~\cite{Maldacena:2002vr,Jeong:2012df,Dai:2013ikl,Dai:2013kra,
Maldacena:2011nz,Gao:2011vs,Gao:2012ib,Sreenath:2013xra}.  
In order to drive a bounce, it is well known that one requires matter 
fields that violate the null energy condition.  
Therefore, analyzing the evolution of the scalar perturbations require 
suitable modelling of the matter fields~\cite{Brandenberger:2012,
Battefeld:2014uga,Novello:2008ra}.
In contrast, the tensor perturbations depend only on the scale factor and 
hence are simpler to study.
For this reason, we shall focus on the tensor bi-spectrum in this work.
Further, we shall assume a specific functional form for the scale factor 
and we shall not attempt to construct sources that can give rise to such 
a behavior.

\par

An interesting aspect of the three-point functions is their property in the 
so-called squeezed limit wherein the wavelength of one of the three modes 
involved is much larger than the other 
two~\cite{Maldacena:2002vr,Creminelli:2004yq,Cheung:2007sv,RenauxPetel:2010ty,
Ganc:2010ff,Creminelli:2011rh,Martin:2012pe,Sreenath:2014nca,Kundu:2014gxa,
Kundu:2015xta}.
In this limit, under fairly general conditions, it is known that the three-point 
functions can be expressed completely in terms of the two-point functions, a 
relation that is referred to as the consistency condition.
We should mention that, while the scalar consistency relation has drawn most 
of the attention, it has been established that all the four three-point 
functions involving scalars and tensors satisfy similar relations under certain 
conditions~\cite{Jeong:2012df,Dai:2013ikl,Dai:2013kra,Kundu:2013gha,Sreenath:2014nka}. 
It is interesting to examine if the three-point functions generated in the
bouncing scenarios satisfy the consistency condition.
In the context of inflation, it is well known that the consistency relations 
arise due to the fact that the amplitude of the long wavelength mode freezes
on super-Hubble scales.
In contrast, in a bouncing universe it can be readily shown that the amplitude
of the long wavelength mode grows sharply as one approaches the bounce during 
the contracting phase.
This behavior suggests that the consistency relation may not hold in bouncing
models~\cite{Cai:2009fn}.
The primordial consistency conditions lead to corresponding
imprints on the anisotropies in the cosmic microwave background (see, for
instance, Refs.~\cite{Creminelli:2011sq,Lewis:2012tc,Shiraishi:2013vja}; in 
particular, see Ref.~\cite{Kehagias:2014caa} for the signatures of the tensor 
modes) and the large scale structure (see, for example, 
Refs.~\cite{Liguori:2010hx,Chiang:2014oga,Mirbabayi:2014gda}).
It is clear that the consistency condition, if it can be confirmed by the 
observations, can help us discriminate between models of the early universe.

\par

The most comprehensive formalism to study the generation of non-Gaussianities
in the early universe is the approach due to Maldacena~\cite{Maldacena:2002vr}.
In this work, we analytically evaluate the tensor bi-spectrum in a matter
bounce using the Maldacena formalism.
To arrive at the tensor bi-spectrum analytically, one requires not only the 
behavior of the tensor modes, one also needs to be able to evaluate a certain 
integral involving the scale factor and the tensor modes. 
We conveniently divide the evolution into three domains and use the analytic 
solutions available in these domains to carry out the integrals and obtain the 
tensor bi-spectrum.

\par

This paper is organized as follows.
In the following section, considering a specific form for the scale factor, 
we shall divide the period before the bounce into two domains, the first 
corresponding to early times during the contracting phase and the other 
close to the bounce.
We shall describe the analytic solutions to the tensor modes during these 
two domains and also discuss the behavior of the modes after the bounce to
arrive at the corresponding tensor power spectrum over wavenumbers much 
smaller than the wavenumber associated with the bounce.
We shall also compare the analytical solutions for the tensor modes with
the corresponding results obtained numerically. 
In Sec.~\ref{sec:tbs-hnl}, we shall quickly summarize the essential 
expressions describing the tensor bi-spectrum in the Maldacena formalism.
We shall also introduce the tensor non-Gaussianity parameter $\hnl$, a 
dimensionless quantity that reflects the amplitude of the tensor 
bi-spectrum.
In Sec.~\ref{sec:ev-tbs}, we shall evaluate the tensor bi-spectrum using the 
analytic solutions to the modes and the behavior of the scale factor in 
the three domains.
We shall calculate the bi-spectrum for an arbitrary triangular configuration 
of the wavevectors.
In Sec.~\ref{sec:r}, we shall illustrate the results in the equilateral and 
the squeezed limits.
We shall show that the non-Gaussianity parameter $\hnl$ that characterizes 
the tensor bi-spectrum is very small for cosmological scales and that the 
consistency relation is violated in the squeezed limit.
We shall conclude with a brief discussion in Sec.~\ref{sec:d}.
In an Appendix, we shall briefly outline a proof of the consistency
condition satisfied by the tensor bi-spectrum during inflation.

\par

Note that we shall work with natural units wherein $\hbar=c=1$, and define 
the Planck mass to be $\Mpl=(8\,\pi\, G)^{-1/2}$.


\section{The tensor modes and the power spectrum}\label{sec:tm-ps}

We shall consider the background to be the spatially 
flat Friedmann-Lema\^itre-Robertson-Walker (FLRW) metric that is 
described by the line-element
\begin{equation}
\d s^2=a^2(\eta)\,\l(-\d \eta^2+\delta_{ij}\,\d {\bm x}^i\,
\d {\bm x}^j\r), 
\end{equation}
where $a(\eta)$ denotes the scale factor and $\eta$ is the conformal 
time coordinate.
We shall assume that the scale factor describing the bouncing scenario is
given in terms of the conformal time coordinate $\eta$ by the relation
\begin{equation}
a(\eta) = a_0\,\l(1 + \eta^2/\eta_0^2\r)
=a_0\,\l(1 + k_0^2\,\eta^2\r),\label{eq:sf}
\end{equation}
where $a_0$ is the minimum value of the scale factor at the bounce (\ie when 
$\eta = 0$) and $\eta_0=k_0^{-1}$ denotes the time scale that determines the 
duration of the bounce. 
Note that, at very early times, \viz when $\eta\ll -\eta_0$, the scale factor 
behaves as in a matter dominated universe (\ie as $a \propto \eta^2$) and, 
for this reason, such a bouncing model is often referred to as the matter bounce.
We shall assume that the scale associated with the bounce, \viz $k_0$, is of 
the order of the Planck scale~$\Mpl$.
Therefore, the wavenumbers of cosmological interest are $50$--$60$ orders of 
magnitude smaller than the wavenumber $k_0$. 

\par

Upon taking into account the tensor perturbations characterized by $\g_{ij}$, 
the spatially flat FLRW metric can be expressed as~\cite{Maldacena:2002vr}
\begin{equation}
\d s^2 =a^{2}(\eta)\; \l(-\d \eta^2 
+ \l[{\rm e}^{\gamma(\eta,{\bm x})}\r]_{ij}
\d {\bm x}^i\, \d {\bm x}^j\r).\label{eq:metric}
\end{equation}
Recall that the primordial perturbations are generated due to quantum 
fluctuations.
On quantization, the tensor perturbation $\hat\g_{ij}$ can be decomposed 
in terms of the corresponding Fourier modes $h_k$ as follows:
\begin{eqnarray}
\hat{\gamma}_{ij}(\eta, {\bf x}) 
&=& \int \frac{\d^{3}{\bm k}}{\l(2\,\pi\r)^{3/2}}\,
\hat{\gamma}_{ij}^{\bm k}(\eta)\, {\rm e}^{i\,{\bm k}\cdot{\bm x}}\nn\\
&=& \sum_{s}\int \frac{\d^{3}{\bm k}}{(2\,\pi)^{3/2}}\,
\l(\hat{a}^{s}_{\bm k}\, \varepsilon^{s}_{ij}({\bm k})\,
h_{k}(\eta)\, {\rm e}^{i\,{\bm k}\cdot{\bm x}}
+\hat{a}^{s\dagger}_{\bf k}\,\varepsilon^{s*}_{ij}({\bm k})\, h^{*}_{k}(\eta)\,
{\rm e}^{-i\,{\bm k}\cdot{\bm x}}\r).\label{eq:t-m-dc}
\end{eqnarray}
In this decomposition, the pair of operators $(\hat{a}_{\bm k}^{s},
\hat{a}^{s\dagger}_{\bm k})$ represent the annihilation and creation 
operators corresponding to the tensor modes associated with the 
wavevector ${\bm k}$, and they satisfy the standard commutation relations.
The quantity $\varepsilon^{s}_{ij}({\bm k})$ represents the polarization 
tensor of the gravitational waves with their helicity being denoted by the 
index~$s$.
The transverse and traceless nature of the gravitational waves leads to
the conditions $\varepsilon^{s}_{ii}({\bm k})=k_{i}\,
\varepsilon_{ij}^s({\bm k})=0$. 
We shall work with a normalization such that $\varepsilon_{ij}^{r}({\bm k})\,
\varepsilon_{ij}^{s*}({\bm k})=2\;\delta^{rs}$~\cite{Maldacena:2002vr}.
The tensor power spectrum, \viz\/ ${\mathcal P}_{_{\rm T}}(k)$, is defined 
as follows:
\begin{eqnarray}
\langle\, {\hat \gamma}_{ij}^{\bm k}(\ee)\,
{\hat \gamma}_{mn}^{\bm k'}(\ee)\,\rangle
&=&\f{(2\,\pi)^2}{2\, k^3}\, \f{\Pi_{ij,mn}^{{\bm k}}}{4}\;
{\mathcal P}_{_{\rm T}}(k)\;
\delta^{(3)}({\bm k}+{\bm k'}),\label{eq:tps-d}
\end{eqnarray}
where the expectation values on the left hand sides are to 
be evaluated in the specified initial quantum state of the 
perturbations, and $\ee$ denotes a suitably late conformal
time when the power spectrum is to be evaluated.
The quantity $\Pi_{ij,mn}^{\vk}$ is given 
by~\cite{Jeong:2012df,Dai:2013ikl,Dai:2013kra,Kundu:2013gha,Sreenath:2014nka}
\begin{equation}
\Pi_{ij,mn}^{\vk}
=\sum_{s}\;\varepsilon_{ij}^{s}(\vk)\;
\varepsilon_{mn}^{s\ast}(\vk).
\end{equation}
The tensor spectral index $\nt$ is defined as
\begin{equation}
\nt=\f{\d\, {\rm ln}\,\pt(k)}{\d\, {\rm ln}\, k}. 
\end{equation}

\par

The tensor modes $h_k$ satisfy the differential equation
\begin{equation}
h_k''+2\,\frac{a'}{a}\,h_k'+k^2\,h_k=0,\label{eq:de-hk}
\end{equation}
where the overprimes denote differentiation with respect to the 
conformal time $\eta$.
If we write $h_k=u_k/a$, then the modes $u_k$ satisfy the equation 
\begin{equation}
u_k''+\l(k^2-\f{a''}{a}\r)\,u_k=0. 
\end{equation}
The quantity $a''/a$ corresponding to the scale factor~(\ref{eq:sf}) is 
given by 
\begin{equation}
\f{a''}{a}=\f{2\,k_0^2}{1+k_0^2\,\eta^2},
\end{equation}
which has essentially a Lorentzian profile.
Note that the quantity $a''/a$ exhibits a maximum at the bounce, with the 
maximum value being of the order of $k_0^2$.
Also, it goes to zero as $\eta\to\pm \infty$. 
For modes of cosmological interest, one finds that $k^2\gg a''/a$ at suitably
early times (\ie as $\eta\to-\infty$).
In this domain, the quantity $u_k$ oscillates and we can impose the standard 
initial conditions on these modes and study their evolution thereafter.

\par

Let us divide the period before the bounce into two domains, a domain 
corresponding to early times and another closer to the bounce.
Let the first domain be determined by the condition $-\infty<\eta < -\alpha\,
\eta_0$, where $\alpha$ is a relatively large number, which we shall set to 
be, say, $10^5$.
The second domain evidently corresponds to $-\alpha\,\eta_0<\eta<0$.
In the first domain, we can assume that the scale factor behaves as
\begin{equation}
a(\eta)\simeq a_0\, k_0^2\, \eta^2,\label{eq:sf-d1}
\end{equation}
so that $a''/a=2/\eta^2$.
Since the condition $k^2=a''/a$ corresponds to, say, $\eta_k=-\sqrt{2}/k$,
the initial conditions can be imposed when $\eta\ll \eta_k$. 
The modes $h_k$ can be easily obtained in such a case and the positive
frequency modes that correspond to the vacuum state at early times are 
given by~\cite{Starobinsky:1979ty,Wands:1998yp,Finelli:2001sr}
\begin{equation}
h_k=\f{\sqrt{2}}{\Mpl}\,\f{1}{\sqrt{2\,k}}\,\f{1}{a_0\,k_0^2\,\eta^2}\,
\l(1-\f{i}{k\,\eta}\r)\,{\rm e}^{-i\,k\,\eta}.\label{eq:hk-d1}
\end{equation}

\par

Let us now consider the behavior of the modes in the second domain, \ie
when $-\alpha\,\eta_0<\eta<0$.
Since we are interested in scales much smaller than $k_0$, we shall assume 
that $\eta_k \ll -\alpha\,\eta_0$, which corresponds to the condition $k\ll 
k_0/\alpha$.
Therefore, in this domain, for scales of cosmological interest, the equation 
governing the tensor mode $h_k$ reduces to  
\begin{equation}
h_k''+\frac{2\,a'}{a}\,h_k'\simeq 0.
\end{equation}
This equation can be immediately integrated to yield
\begin{equation}\label{eq:h_eta_prime}
h_k'(\eta)\simeq h_k'(\eta_\ast)\,\f{a^2(\eta_\ast)}{a^2(\eta)},
\end{equation}
where $\eta_\ast$ is a suitably chosen time and the scale factor $a(\eta)$ 
is given by the complete expression~(\ref{eq:sf}).
On further integration, we obtain that
\begin{equation}
h_k(\eta)
=h_k(\eta_\ast)+h_k'(\eta_\ast)\,a^2(\eta_\ast)\,
\int_{\eta_\ast}^{\eta} \frac{{\rm d}{\tilde \eta}}{a^2({\tilde \eta})},
\label{eq:hk-shs}
\end{equation}
where we have chosen the constant of integration to be $h_k(\eta_\ast)$.
If we choose $\eta_\ast=-\alpha\,\eta_0$, we can make use of the 
solution~(\ref{eq:hk-d1}) to determine $h_k(\eta_\ast)$ and 
$h_k'(\eta_\ast)$.
Note that, in the domain of interest, the first term in the above expression
is, evidently, a constant, while the second term grows rapidly as one 
approaches the bounce.
Upon using the form~(\ref{eq:sf}) of the scale factor, we find that we can 
express the behavior of the mode $h_k$ in the second domain as
\begin{equation}
h_k= A_k+B_k\,f(k_0\,\eta),\label{eq:hk-d2}
\end{equation}
where 
\begin{equation}
f(k_0\,\eta)=\f{k_0\,\eta}{1+k_0^2\,\eta^2}+\tan^{-1}\l(k_0\,\eta\r),
\label{eq:f}
\end{equation}
while the quantities $A_k$ and $B_k$ are given by
\begin{eqnarray}
A_k &=& \f{\sqrt{2}}{\Mpl}\,\f{1}{\sqrt{2\,k}}\,\frac{1}{a_0\,\alpha^2}\,
\l(1+\frac{i\,k_0}{\alpha\,k}\right)\,{\rm e}^{i\,\alpha\,k/k_0}
+ B_k\, f(\alpha),\\
B_k &=& \f{\sqrt{2}}{\Mpl}\,\f{1}{\sqrt{2\,k}}\,\frac{1}{2\,a_0\,\alpha^2}\,
\l(1+\alpha^2\r)^2\,\l(\f{3\,i\,k_0}{\alpha^2\,k}+\frac{3}{\alpha}
- \frac{i\,k}{k_0}\r)\,{\rm e}^{i\,\alpha\,k/k_0}.
\end{eqnarray}

\par

Let us now turn to the third domain, \ie immediately after the bounce.
In this case too, for modes such that $k\ll k_0/\alpha$, the solution to 
$h_k$ is given by Eq.~(\ref{eq:hk-d2}).
We should highlight the fact that, whereas the bounce~(\ref{eq:sf}) is a 
symmetric one, the solution~(\ref{eq:hk-d2}) is asymmetric in $\eta$.
Moreover, one may have naively expected the amplitude of the long wavelength 
modes to freeze once the universe starts expanding.  
This is largely true, though not completely so, and the behavior can possibly
be attributed to the specific form of the scale factor~(\ref{eq:sf}). 
Note that, during this domain, while the first term in $f(k_0\,\eta)$ 
decays, the second term actually grows, albeit extremely mildly. 
We shall assume that, after the bounce, the universe transits to the
conventional radiation domination epoch at, say, $\eta=\beta\,\eta_0$,
where we shall set $\beta\simeq 10^2$.
We should mention that this choice is somewhat arbitrary and we shall discuss 
the dependence of the tensor power spectrum and the bi-spectrum on $\beta$ in 
due course.

\par

In order to understand the extent of accuracy of the approximations involved, 
it would be worthwhile to compare the above analytical results for the mode 
$h_k$ with the corresponding numerical results. 
Clearly, given the scale factor~(\ref{eq:sf}), it is a matter of integrating
the differential equation~(\ref{eq:de-hk}), along with the standard Bunch-Davies
initial conditions, to arrive at the behavior of $h_k$.
The conformal time coordinate does not prove to be an efficient time variable
for numerical integration, particularly when a large range in the scale factor
needs to be covered.
In the context of inflation, it is often the e-fold $N$, defined as $a(N) =a_0\, 
{\rm exp}\,N$, where $N=0$ is a suitable time at which the scale factor takes the 
value $a_0$, that is utilized to integrate the equations of motion (see, for 
instance, Refs.~\cite{Salopek:1988qh,Ringeval:2007am,Jain:2008dw,Hazra:2012yn}).
Due to the exponential factor involved, a small range in e-folds covers a 
large range in time and scale factor.
However, since ${\rm e}^{N}$ is a monotonically increasing function, while
it is useful to describe expanding universes, e-folds are not helpful in 
characterizing bouncing scenarios.   
In order to characterize a bounce, it would be convenient to choose a variable 
that is negative during the contracting phase of the universe, zero at the 
bounce and positive during the expanding phase. 
We shall choose to perform the integration using a new variable $\cN$, which 
we call the e-N-fold, in terms of which the scale factor is defined as $a(\cN) 
= a_0\, {\rm exp}\l(\cN^2/2\r)$~\cite{Sriramkumar:2015yza}.
We shall assume that $\cN$ is zero at the bounce, with negative values 
representing the phase prior to the bounce and positive values after.

\par

\begin{figure}[!t]
\begin{center}
\includegraphics[width=12.00cm]{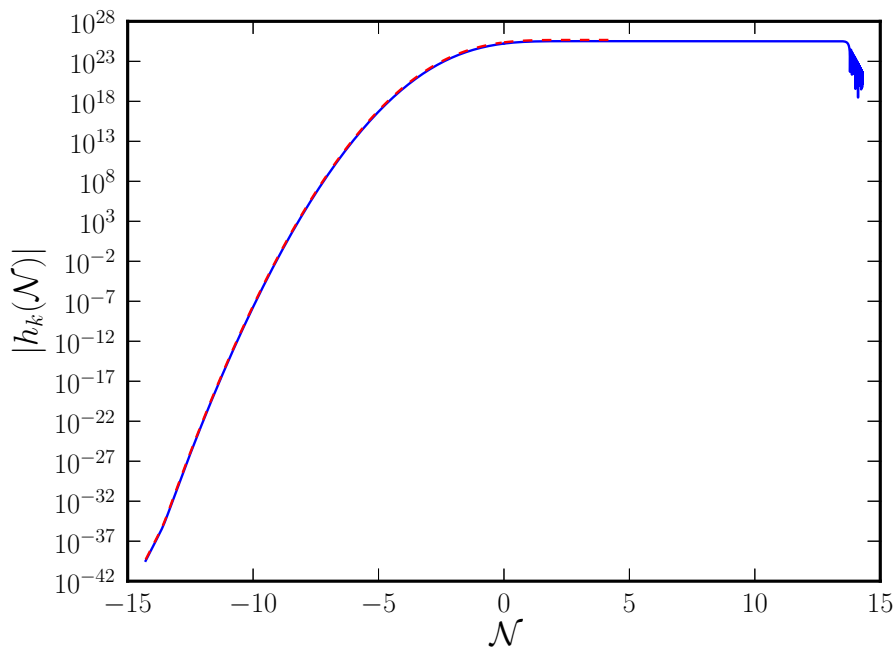}
\end{center}
\caption{A comparison of the numerical results (in blue) with the analytical
results (in red) for the amplitude of the tensor mode $\vert h_k\vert$ 
corresponding to the wavenumber $k/k_0 = 10^{-20}$.  
We have set $a_0=10^5$ and, for plotting the analytical results, we have chosen 
$\alpha=10^5$.
We have plotted the results from the initial e-N-fold $\cN_i$ [when $k^2
=10^4\, (a''/a)$] corresponding to the mode.
While we have illustrated the exact numerical result till rather late times,
we have plotted the analytical results until a time after the bounce when the 
power spectrum is evaluated (see discussion below). 
Evidently, the analytical and numerical results match extremely well, suggesting 
that the analytical approximation for the modes works to a very good accuracy.}
\label{fig:hk}
\end{figure}

In terms of the e-N-fold, the differential equation~(\ref{eq:de-hk}) 
governing the evolution of the tensor modes can be expressed as
\begin{equation}
\f{\d^2 h_k}{\d\cN^2}
+\l(3\,\cN+\f{1}{H}\,\f{\d H}{\d\cN}-\f{1}{\cN}\r)\,\f{\d h_k}{\d\cN}
+\l(\f{k\,\cN}{a\,H}\r)^2\,h_k=0,\label{eq:de-hk-cN}
\end{equation}
where $H=a'/a^2$ is the Hubble parameter.
Given the scale factor~(\ref{eq:sf}), the corresponding Hubble parameter 
can be easily evaluated in terms of the conformal time $\eta$. 
In order to express the Hubble parameter $H$ in terms of the e-N-fold, 
we shall require $\eta$ as a function of~$\cN$. 
Upon using the definition of the e-N-folds and the expression~(\ref{eq:sf})
for the scale factor, we obtain that
\begin{equation}
\eta(\cN)= \pm\, k_0^{-1}\, \l({\rm e}^{\cN^2/2}-1\r)^{1/2}.
\end{equation}
Since the Hubble parameter is negative during the contracting phase and 
positive during the expanding regime, we have to choose the root of 
$\eta(\cN)$ accordingly during each phase.
From the expression for the Hubble parameter, we evaluate the coefficients 
of the differential equation~(\ref{eq:de-hk-cN}) in terms of $\cN$.
With the coefficients at hand, we numerically integrate the differential
equation using a fifth order Runge-Kutta algorithm.
We should mention that we have also independently checked the numerical 
results using {\it Mathematica}.\/
Recall that, the initial conditions are imposed in a domain during the 
contracting phase wherein $k^2\gg a^{\prime\prime}/a$.
As is done in the context of inflation, we shall impose the initial conditions 
when $k^2=10^4\, (a^{\prime\prime}/a)$ corresponding to, say, the e-N-fold 
$\cN_i$.
It should be pointed out that the initial conditions on the different modes
are imposed at different times.
In terms of the e-N-folds, the standard Bunch-Davies initial conditions can
be expressed as
\begin{subequations}
\begin{eqnarray}
h_k &=& \f{1}{a(\cN_i)\sqrt{2k}},\\
\f{\d h_k}{\d\cN} 
&=& -\f{i\,\cN_i}{a^2(\cN_i)\,H(\cN_i)}\,\sqrt{\f{k}{2}}
-\f{\cN_i}{a(\cN_i)\,\sqrt{2k}}.
\end{eqnarray}
\end{subequations}
We impose these initial conditions well before the bounce and evolve the 
modes until a suitable time after the bounce.
The tensor mode $h_k$ evaluated in such a manner has been plotted for a 
given wavenumber (such that $k/k_0\ll 1$) in Fig.~\ref{fig:hk}. 
The figure also contains a plot of the analytical results~(\ref{eq:hk-d1}) 
and~(\ref{eq:hk-d2}) for the same wavenumber.
As is evident from the figure, prior to the bounce and immediately after, 
the analytical results match the exact numerical results exceedingly well.

\par

\begin{figure}[!tbp] 
\begin{center}
\includegraphics[width=12.00cm]{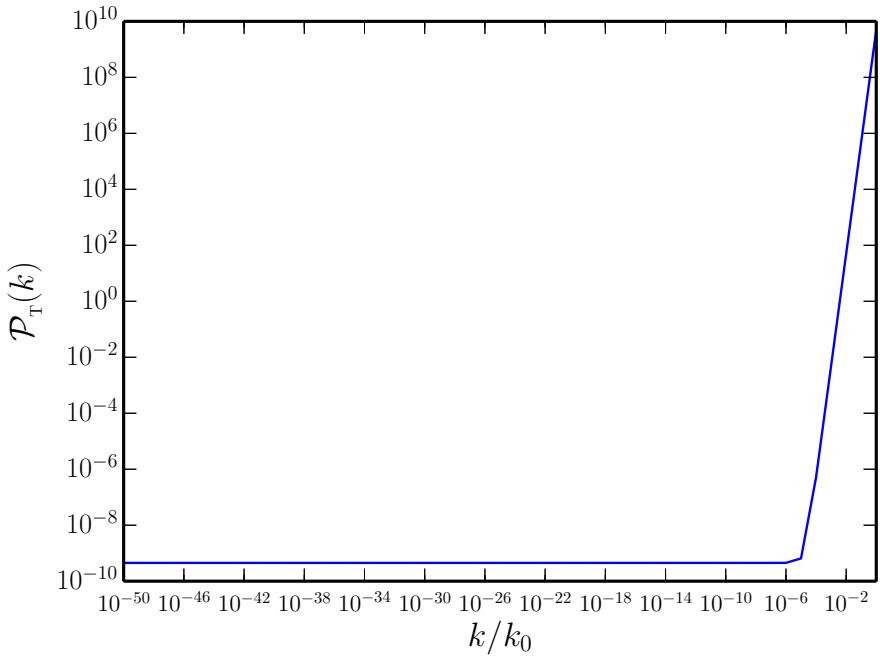}
\end{center}
\vskip -20pt
\caption{The behavior of the tensor power spectrum has been plotted as a 
function of $k/k_0$ for a wide range of wavenumbers. 
In plotting this figure, we have set $k_0/\Mpl=1$, $a_0=10^5$, $\alpha=10^5$
and $\beta=10^2$.
We should emphasize that the approximations we have worked with are valid
only over the domain wherein $k\ll k_0/\alpha$.
Clearly, the power spectrum is scale invariant in this domain.
We also find that, at small wavenumbers, the tensor power spectrum has 
the expected scale invariant amplitude of ${\cal P}_{_{\rm T}}(k)=4.5\times 
10^{-10}$ corresponding to $k_0/\Mpl=1$ and $a_0=10^5$.}
\label{fig:tps}
\end{figure}

The tensor power spectrum after the bounce can be calculated using the 
solutions we have obtained.
Recall that the tensor power spectrum is defined as 
\begin{equation}
{\cal P}_{_{\rm T}}(k) = 4\,\frac{k^3}{2\,\pi^2}\,\vert h_k(\eta)\vert^2,
\end{equation}
with the spectrum to be evaluated at a suitable time.
If we evaluate the tensor power spectrum at $\eta=\beta\,\eta_0$, we find
that it can be expressed as
\begin{equation}
{\cal P}_{_{\rm T}}(k) 
= 4\,\frac{k^3}{2\,\pi^2}\,\vert A_k+B_k\, f(\beta)\vert^2.\label{eq:tps}
\end{equation}
Note that, $\alpha$ is a quantity that we have artificially introduced and
the actual problem does not contain $\alpha$. 
For $k\ll k_0/\alpha$ and a sufficiently large $\alpha$ (as we had said, 
for $\alpha=10^5$ or so), the above power spectrum reduces to a scale
invariant form with a weak dependence on $\beta$, if $\beta$ is reasonably 
larger than unity. 
If we further assume that $\beta$ is large enough (say, $10^2$), then the 
scale invariant amplitude is found to be: ${\cal P}_{_{\rm T}}(k)\simeq 
9\,k_0^2/(2\,\Mpl^2\,a_0^2)$, as 
expected~\cite{Starobinsky:1979ty,Wands:1998yp,Finelli:2001sr}.
In Fig.~\ref{fig:tps}, we have plotted the complete tensor power spectrum
described by the expression~(\ref{eq:tps}) for a given set of parameters. 
We should stress that the power spectrum is actually valid only for modes 
which satisfy the condition $k\ll k_0/\alpha$. 
It is evident from the figure that the power spectrum is strictly scale 
invariant over this domain.
Moreover, we find that the spectrum indeed reduces to the above-mentioned 
scale invariant amplitude for small values of the wavenumbers.
We have also evaluated the tensor power spectrum numerically
using the method described above.
We have computed the spectrum at a given time soon after the bounce 
(corresponding to $\beta=10^2$) for all the modes. 
We find that, for wavenumbers such that $k\ll k_0$, the numerical analysis 
also leads to a scale invariant spectrum whose amplitude matches the above 
analytical result to about $1\%$.

\par


\section{The tensor bi-spectrum and the corresponding
non-Gaussianity\\ parameter}\label{sec:tbs-hnl}

As we have mentioned, the most comprehensive formalism to calculate the 
three-point functions generated in the early universe is the formalism 
due to Maldacena~\cite{Maldacena:2002vr}. 
The primary aim of Maldacena's approach is to obtain the cubic order action 
that governs the scalar and the tensor perturbations using the ADM formalism. 
Then, based on the action, one arrives at the corresponding three-point 
functions using the standard rules of perturbative quantum field theory.

\par

The tensor bi-spectrum in Fourier space, \viz\/
$\cB_{\gamma\gamma\gamma}^{m_1n_1m_2n_2m_3 n_3}(\vka,\vkb,\vkc)$, evaluated
at the conformal time, say, $\ee$, is defined as
\begin{eqnarray}
\langle\, {\hat \gamma}_{m_1n_1}^{\vka}(\eta _{\rm e})\,
{\hat \gamma}_{m_2n_2}^{\vkb}(\eta _{\rm e})\, 
{\hat \gamma}_{m_3n_3}^{\vkc}(\eta _{\rm e})\,\rangle 
&\equiv& \l(2\,\pi\r)^3\, \cB_{\g\g\g}^{m_1n_1m_2 n_2 m_3 n_3}(\vka,\vkb,\vkc)\;
\delta^{(3)}\l(\vka+\vkb+\vkc\r).\label{eq:ttt}\nn\\
\end{eqnarray}
Note that the delta function on the right hand side implies that the wavevectors 
$\vka$, $\vkb$ and $\vkc$ form the edges of a triangle.
For convenience, hereafter, we shall set
\begin{equation}
\cB_{\g\g\g}^{m_1n_1m_2n_2m_3n_3}(\vka,\vkb,\vkc)
= \l(2\,\pi\r)^{-9/2}\, G_{\g\g\g}^{m_1n_1m_2n_2m_3n_3}(\vka,\vkb,\vkc).
\end{equation}
The tensor bi-spectrum 
$G_{\gamma\gamma\gamma}^{m_1n_1m_2n_2m_3n_3}(\vka,\vkb,\vkc)$,
calculated in the perturbative vacuum using the Maldacena formalism, can 
be written in terms of the modes $h_k$ 
as follows~\cite{Maldacena:2002vr,Maldacena:2011nz,Gao:2011vs,
Sreenath:2013xra,Kundu:2013gha,Gao:2012ib}:
\begin{eqnarray} \label{eq:Gggg}
G_{\gamma\gamma\gamma}^{m_1n_1m_2n_2m_3n_3}(\vka,\vkb,\vkc)
&= & \Mp^2\; \biggl[\bigl(\Pi_{m_1n_1,ij}^{\vka}\,\Pi_{m_2n_2,im}^{\vkb}\,
\Pi_{m_3n_3,lj}^{\vkc}\nn\\
& &-\,\f{1}{2}\;\Pi_{m_1n_1,ij}^{\vka}\,\Pi_{m_2n_2,ml}^{\vkb}\,
\Pi_{m_3n_3,ij}^{\vkc}\bigr)\, k_{1m}\, k_{1l}
+{\rm five~permutations}\biggr]\nn\\
& &\times\; \bigl[h_{\ska}(\ee)\, h_{\skb}(\ee)\, h_{\skc}(\ee)\,
\cG_{\gamma\gamma\gamma}(\vka,\vkb,\vkc)\nn\\
& &+\, {\rm complex~conjugate}\bigr],\label{eq:tbs-ov}
\end{eqnarray}
where $\cG_{\gamma\gamma\gamma}(\vka,\vkb,\vkc)$ is described by the integral
\begin{equation}
\cG_{\g\g\g}(\vka,\vkb,\vkc)
=-\f{i}{4}\,\int_{\ei}^{\ee} \d\eta\; a^2\, h_{\ska}^{\ast}\,
h_{\skb}^{\ast}\,h_{\skc}^{\ast},\label{eq:cGggg}
\end{equation}
with $\ei$ denoting the time when the initial conditions are imposed on the 
perturbations and $\ee$ representing the time when the bi-spectrum is to be
evaluated.
Also, we should mention that $(k_{1i},k_{2i}, k_{3i})$ denote the 
components of the three wavevectors $({\bm k}_1, {\bm k}_{2}, {\bm k}_{3})$ 
along the $i$-spatial direction\footnote{Such a clarification seems
necessary to avoid confusion between $k_1$, $k_2$ and $k_3$ which
denote the wavenumbers associated with the wavevectors ${\bm k}_1$, 
${\bm k}_{2}$ and ${\bm k}_{3}$, and the quantity $k_i$ which 
represents the component of the wavevector ${\bm k}$ along the 
$i$-spatial direction.}.

\par

The dimensionless non-Gaussianity parameter that characterizes the amplitude 
of the tensor bi-spectrum is defined as~\cite{Sreenath:2013xra} 
\begin{eqnarray}
\hnl(\vka,\vkb,\vkc)
&=&-\l(\f{4}{2\,\pi^2}\r)^2\,
\l[k_1^3\, k_2^3\, k_3^3\;
G_{\g\g\g}^{m_1n_1m_2n_2m_3n_3}(\vka,\vkb,\vkc)\r]\nn\\
&\times&\; \l[\Pi_{m_1n_1,m_3n_3}^{\vka}\,\Pi_{m_2n_2,{\bar m}{\bar n}}^{\vkb}\;
k_3^3\; {\mathcal P}_{_{\rm T}}(k_1)\;{\mathcal P}_{_{\rm T}}(k_2)
+{\rm five~permutations}\r]^{-1},\qquad\label{eq:hnl-ov}
\end{eqnarray}
where the overbars on the indices imply that they need to be summed over all
allowed values.
Our aim in this work is to evaluate the magnitude and shape of the tensor
bi-spectrum and the corresponding non-Gaussianity parameter and compare 
with, say, the results in de Sitter inflation.
Therefore, for simplicity, we shall set the polarization tensor to unity.
In such a case, the expression~(\ref{eq:tbs-ov}) for the tensor bi-spectrum 
above simplifies to
\begin{eqnarray}
G_{\gamma\gamma\gamma}(\vka,\vkb,\vkc)
&=& \Mp^2\,\bigl[h_{\ska}(\ee)\, h_{\skb}(\ee)\,h_{\skc}(\ee)\,
{\bar \cG}_{\gamma\gamma\gamma}(\vka,\vkb,\vkc)\nn\\
& &+\, {\rm complex~conjugate}\bigr],\label{eq:tbs}
\end{eqnarray}
where the quantity ${\bar \cG}_{\g\g\g}(\vka,\vkb,\vkc)$ is described by
the integral
\begin{eqnarray}
{\bar \cG}_{\g\g\g}(\vka,\vkb,\vkc)
&=&-\f{i}{4}\,\l(k_1^2+k_2^2+k_3^2\r)\,\int_{\ei}^{\ee} \d\eta\; 
a^2\, h_{\ska}^{\ast}\,
h_{\skb}^{\ast}\,h_{\skc}^{\ast}.\label{eq:cGggg-fv}
\end{eqnarray}
We shall choose $\ei$ to be an early time during the contracting phase when 
the initial conditions are imposed on the modes (\ie when $k^2\gg a''/a$), 
and $\ee$ to be a suitably late time, say, some time after the bounce, when 
the bi-spectrum is evaluated.
If we ignore the factors involving the polarization tensor, the non-Gaussianity 
parameter $\hnl$ reduces to
\begin{eqnarray}
\hnl(\vka,\vkb,\vkc)\!\!
&=&\!\!-\l(\f{4}{2\,\pi^2}\r)^2\, \l[k_1^3\, k_2^3\, k_3^3\; 
G_{\g\g\g}(\vka,\vkb,\vkc)\r]\nn\\ 
& &\times\,\biggl[2\,k_3^3\; {\mathcal P}_{_{\rm T}}(k_1)\,
{\mathcal P}_{_{\rm T}}(k_2)+\,{\rm two~permutations}\biggr]^{-1}.\label{eq:hnl}
\end{eqnarray}

\section{Evaluating the tensor bi-spectrum}\label{sec:ev-tbs}

With the forms of the scale factor and the mode functions at hand, in 
order to arrive at the tensor bi-spectrum, it is now a matter of 
evaluating the integral~(\ref{eq:cGggg-fv}) in the three domains.

\par 

Let us begin by considering the first domain.
Upon using the behavior~(\ref{eq:sf-d1}) of the scale factor and the 
mode~(\ref{eq:hk-d1}) in the first domain, we find that the quantity 
${\bar \cG}_{\g\g\g}(\vka,\vkb,\vkc)$ can be expressed as
\begin{eqnarray}
{\bar \cG}^1_{\g\g\g}(\vka,\vkb,\vkc)
&=&\f{-i\,\l(k_1^2+k_2^2+k_3^2\r)}{4\,\Mp^3\,a_0\,k_0^2\,\sqrt{k_1\,k_2\,k_3}}\,
\biggl[I_2(\kT,k_0,\alpha)+\,i\,\l(\frac{1}{k_1}+\frac{1}{k_2}+\f{1}{k_3}\r)\,
I_3(\kT,k_0,\alpha)\nn\\
& &-\,\l(\f{1}{k_1\,k_2}+\frac{1}{k_2\,k_3}
+\frac{1}{k_1\,k_3}\r)\,I_4(\kT,k_0,\alpha)
-\,\f{i}{k_1\,k_2\,k_3}\,I_5(\kT,k_0,\alpha)\biggr], 
\end{eqnarray}
where $\kT=k_1\,+\,k_2\,+\,k_3$ and the quantities $I_n(\kT,k_0,\alpha)$ are
described by the integrals
\begin{eqnarray}
I_n(\kT,k_0,\alpha)
=\int_{-\infty}^{-\alpha/k_0}\f{\d\eta}{\eta^n}\,{\rm e}^{i\,\kT\,\eta}.
\end{eqnarray}
For $n>1$, these integrals can be evaluated to yield
\begin{eqnarray}
I_{n+1}(\kT,k_0,\alpha)
&=& -\f{1}{n}\,\l(-\frac{k_0}{\alpha}\r)^n\,{\rm e}^{-i\,\alpha\,\kT/k_0}
+\,\frac{i\,\kT}{n}\,I_n(\kT,k_0,\alpha),
\end{eqnarray}
while $I_1(\kT,k_0,\alpha)$ is given by 
\begin{equation}
I_1(\kT,k_0,\alpha)=i\,\pi+{\rm Ei}(-i\,\alpha\,\kT/k_0),
\end{equation}
where ${\rm Ei}(x)$ is the exponential integral function~\cite{Gradshteyn:2007}.

\par

Let us now turn to evaluating ${\bar \cG}_{\g\g\g}(\vka,\vkb,\vkc)$ in the 
second domain. 
Upon using the behavior~(\ref{eq:sf}) of the scale factor and the 
mode~(\ref{eq:hk-d2}), we find that the quantity can be expressed as 
\begin{eqnarray}
{\bar \cG}^2_{\g\g\g}(\vka,\vkb,\vkc)
&=&-\frac{i\,a_0^2\,\left(k_1^2\,+\,k_2^2\,+\,k_3^2\right)}{4\,k_0}\,
\biggl[A_{k_1}^\ast\,A_{k_2}^\ast\,A_{k_3}^\ast\,J_0(\alpha) \nn\\
& & +\l(A_{k_1}^\ast\,A_{k_2}^\ast\,B_{k_3}^\ast
+ A_{k_1}^\ast\, B_{k_2}^\ast\,A_{k_3}^\ast
+ B_{k_1}^\ast\, A_{k_2}^\ast\, A_{k_3}^\ast\r)\, J_1(\alpha)\nn\\
& & +\,\l(A_{k_1}^\ast\,B_{k_2}^\ast\,B_{k_3}^\ast
+ B_{k_1}^\ast\, A_{k_2}^\ast\,B_{k_3}^\ast
+ B_{k_1}^\ast\, B_{k_2}^\ast\, A_{k_3}^\ast\r)\, J_2(\alpha) \nn\\
& & +B_{k_1}^\ast\,B_{k_2}^\ast\,B_{k_3}^\ast\,J_3(\alpha)\biggr],
\label{eq:cG-d2}
\end{eqnarray}
where $J_n(\alpha)$ are described by the integrals
\begin{equation}
J_n(\alpha)
=\int_{-\alpha}^0\,\d x\,\l(1+x^2\r)^2\, f^n(x),
\end{equation}
with the function $f(x)$ being given by Eq.~(\ref{eq:f}). 
The integrals $J_0(\alpha)$ and $J_1(\alpha)$ can be readily evaluated
to obtain that
\begin{equation}
J_0(\alpha)=\alpha+\frac{2\,\alpha^3}{3}\,+\,\frac{\alpha^5}{5}
\end{equation}
and
\begin{eqnarray}
J_1(\alpha)
&=&-\frac{1}{2}\,\l(\alpha^2 + \frac{\alpha^4}{2}\right)\,
-\frac{11}{60}+\frac{2\,\l(1+\alpha^2\r)}{15}
+\frac{\l(1+\alpha^2\r)^2}{20}
-\frac{8\,\alpha}{15}\,\tan^{-1}\alpha \nn\\
& &-\,\frac{4\,\alpha}{15}\,\l(1+\alpha^2\r)\,\tan^{-1}\alpha
-\frac{\alpha}{5}\,\l(1+\alpha^2\r)^2\,\tan^{-1}\alpha
+\frac{4}{15}\,{\rm ln}\l(1+\alpha^2\r).
\end{eqnarray}

\par

In contrast, the integrals $J_2(\alpha)$ and $J_3(\alpha)$ are more involved.
The integral $J_2(\alpha)$ can be divided into three parts and written as
\begin{equation}
J_2(\alpha)=J_{21}(\alpha)+J_{22}(\alpha)+J_{23}(\alpha),
\end{equation}
where the integrals $J_{21}(\alpha)$ and $J_{22}(\alpha)$ can be easily 
evaluated to be
\begin{eqnarray}
J_{21}(\alpha)
&=&\int_{-\alpha}^0{\rm d}x\,x^2= \frac{\alpha^3}{3},\\
J_{22}(\alpha)
&=& 2\,\int_{-\alpha}^0{\rm d}x\,x\,
\l(1+x^2\r)\,\tan^{-1}x\nn\\
&=& \alpha^2\,\l(1+\frac{\alpha^2}{2}\r)\,\tan^{-1}\alpha-\frac{1}{2}\,
\l(\alpha-\tan^{-1}\alpha\r)-\frac{\alpha^3}{6}.
\end{eqnarray}
The quantity $J_{23}(\alpha)$ is given by
\begin{equation}
J_{23}(\alpha)
=\int_{-\alpha}^0{\rm d}x\, \l(1+x^2\r)^2\,\l(\tan^{-1}x\r)^2,
\end{equation}
and, upon setting $\tan^{-1}x=y$, it reduces to
\begin{equation}
J_{23}(\alpha)
=\int_{-\tan^{-1}\alpha}^0\d y\,y^2\,\sec^6y.
\end{equation}
The integral involved can be evaluated to be (see, for instance,
Ref.~\cite{Gradshteyn:2007})
\begin{eqnarray}
\int \d y\, y^2\,\sec^6\,y
&=& \frac{-y\,\l(\cos y-2\,y\,\sin y\r)}{10\,\cos^5 y}
-\frac{4\,y\,\l(\cos y-y\,\sin y\r)}{15\,\cos^3 y}+\l(\frac{11}{30}+
\frac{8\,y^2}{15}\r)\,\tan y \nn\\
& &+\,\frac{\tan^3 y}{30}
+\frac{16}{15}\,\sum_{n=1}^{\infty}
\f{(-1)^n\,2^{2\,n}\l(2^{2\,n}-1\r)\,
y^{2\,n+1}}{\l(2\,n+1\r)\,\l(2\,n\r)!}\,B_{2n},\label{eq:s1}
\end{eqnarray}
where $B_{2n}$ are the Bernoulli numbers.
Needless to add, this result can be used to arrive at $J_{23}(\alpha)$.
We should add that the infinite series in the above expression is
convergent, and we find that it can be expressed as 
follows~\cite{Mathematica8.0}:
\begin{eqnarray}
\sum_{n=1}^{\infty}
\f{(-1)^n\,2^{2\,n}\l(2^{2\,n}-1\r)\,
y^{2\,n+1}}{\l(2\,n+1\r)\,\l(2\,n\r)!}\,B_{2n} 
&=& y\,\Biggl\{{\rm ln}\l[\Gamma\l(1+\f{y}{\pi}\r)\r]
+{\rm ln}\l[\Gamma\l(1-\f{y}{\pi}\r)\r]\nn\\
& &-\,{\rm ln}\l[\Gamma\l(1-\f{2y}{\pi}\r)\r]
-{\rm ln}\l[\Gamma\l(1+\f{2y}{\pi}\r)\r]\Biggr\}\nn\\
& &+\,\pi\,\Biggl\{-\zeta^\prime\l(-1,1+\f{y}{\pi}\r)
+\zeta^\prime\l(-1,1-\f{y}{\pi}\r)\nn\\
& &+\,\f{1}{2}\,\zeta^\prime\l(-1,1+\f{2y}{\pi}\r)
-\f{1}{2}\,\zeta^\prime\l(-1,1-\f{2y}{\pi}\r)\Biggr\},\qquad
\end{eqnarray}
where $\zeta^\prime(s,a)$ denotes the derivative of the Hurwitz zeta 
function $\zeta(s,a)$ with respect to the first argument and 
$\Gamma(n)$ is the Gamma function.

\par

Let us now evaluate the last of the integrals, \viz $J_3(\alpha)$. 
It proves to be convenient to divide the integral into four parts as follows:
\begin{equation}
J_3(\alpha)=J_{31}(\alpha)+J_{32}(\alpha)+J_{33}(\alpha)
+J_{34}(\alpha).
\end{equation}
If we set $\tan^{-1}x=y$, we find that the integrals $J_{31}(\alpha)$,
$J_{32}(\alpha)$ and $J_{33}(\alpha)$ can be easily evaluated to be
\begin{eqnarray}
J_{31}(\alpha) 
&=& \int_{-\tan^{-1}\alpha}^0\d y\, \tan^3y
= -\frac{\alpha^2}{2}+\frac{1}{2}\,{\rm ln}\l(1+\alpha^2\r),\\
J_{32}(\alpha) 
&=& 3\,\int_{-\tan^{-1}\alpha}^0 \d y\, y^2\,\tan y\,\sec^4y\nn\\
&=& -\f{3}{4}\,\l(1+\alpha^2\r)^2\,\l(\tan^{-1}\alpha\r)^2
+\f{\alpha}{2}\,\l(1+\alpha^2\r)\,\tan^{-1}\alpha -\f{\alpha^2}{4}\nn\\
& &+\,\alpha\,\tan^{-1}\alpha-\f{1}{2}\,{\rm ln}\l(1+\alpha^2\r),\\
J_{33}(\alpha) 
&=& 3\,\int_{-\tan^{-1}\alpha}^0 \d y\, y\,\tan^2y\,\sec^2y
= \f{\alpha^2}{2}-\alpha^3\,\tan^{-1}\alpha
-\f{1}{2}\,{\rm ln}\l(1+\alpha^2\r).
\end{eqnarray}
The integral $J_{34}(\alpha)$ is given by
\begin{equation}
J_{34}(\alpha)
=\int_{-\tan^{-1}\alpha}^0\d y\,y^3\,\sec^6y,
\end{equation}
which can be evaluated using the result~\cite{Gradshteyn:2007}
\begin{eqnarray}
\int \d y\, y^3\,\sec^6y
&=& -\f{y^2\,\l(3\,\cos y-4\,y\,\sin y\r)}{20\,\cos^5y}\,
-\f{2\,y^2\,\l(3\,\cos y-2\,y\,\sin y\r)}{15\,\cos^3y} \nn\\
& & +\l(y+\f{8\,y^3}{15}\r)\,\tan y+ {\rm ln}\l|\cos y\r|
+\,\f{y\,\sin y}{10\,\cos^3y}-\frac{1}{20\,\cos^2y} \nn\\
& & +\frac{8}{5}\,\sum_{n=1}^{\infty}\f{(-1)^n\,2^{2\,n}\,\l(2^{2\,n}\,-1\r)\,
y^{2\,n+2}}{(2\,n+2)\,(2\,n)!}\, B_{2n}.\label{eq:s2}
\end{eqnarray}
The infinite series in the above expression is convergent, and it can 
be expressed as~\cite{Mathematica8.0}
\begin{eqnarray}
\sum_{n=1}^{\infty}\f{(-1)^n\,2^{2\,n}\,\l(2^{2\,n}\,-1\r)\,
y^{2\,n+2}}{(2\,n+2)\,(2\,n)!}\, B_{2n} 
&=& y^2\,\Biggl\{{\rm ln}\l[\Gamma\l(1+\f{y}{\pi}\r)\r]
+{\rm ln}\l[\Gamma\l(1-\f{y}{\pi}\r)\r]\nn\\
& &-\,{\rm ln}\l[\Gamma\l(1-\f{2y}{\pi}\r)\r]
-{\rm ln}\l[\Gamma\l(1+\f{2y}{\pi}\r)\r]\Biggr\}+\f{3}{8}\,\zeta(3)\nn\\
& &+\,\pi^2\,\Biggl[\zeta^\prime\l(-2,1+\f{y}{\pi}\r)
+\zeta^\prime\l(-2,1-\f{y}{\pi}\r)\nn\\
& &-\,\f{1}{4}\,\zeta^\prime\l(-2,1-\f{2y}{\pi}\r)
-\f{1}{4}\,\zeta^\prime\l(-2,1+\f{2y}{\pi}\r)\Biggr]\nn\\
& &+\,\pi\, y\,\Biggl[\zeta^\prime\l(-1,1+\f{2y}{\pi}\r)
-\zeta^\prime\l(-1,1-\f{2y}{\pi}\r)\nn\\
& &+\,2\,\zeta^\prime\l(-1,1-\f{y}{\pi}\r)
-2\,\zeta^\prime\l(-1,1+\f{y}{\pi}\r)\Biggr],
\end{eqnarray}
where, as before, $\zeta(s)$ is the Riemann zeta function, $\zeta^\prime(s,a)$ 
denotes the derivative of the Hurwitz zeta function $\zeta(s,a)$ with respect 
to the first argument and $\Gamma(n)$ is the Gamma function.

\par

Let us now consider the quantity ${\bar \cG}_{\g\g\g}(\vka,\vkb,\vkc)$ in the 
third domain, \ie from the bounce at $\eta=0$ to $\eta=\beta\,\eta_0$.
In this domain, the modes $h_k$ and the scale factor have the same form 
as in the second domain.
Therefore, it should be clear that, the quantity 
${\bar \cG}_{\g\g\g}^3(\vka,\vkb,\vkc)$ too will be given by the 
expression~(\ref{eq:cG-d2}), but with the integrals $J_n(\alpha)$ being
replaced by $-J_n(-\beta)$.

\par


\section{Results}\label{sec:r}

\begin{figure}[!htbp] 
\begin{center}
\includegraphics[width=12.00cm]{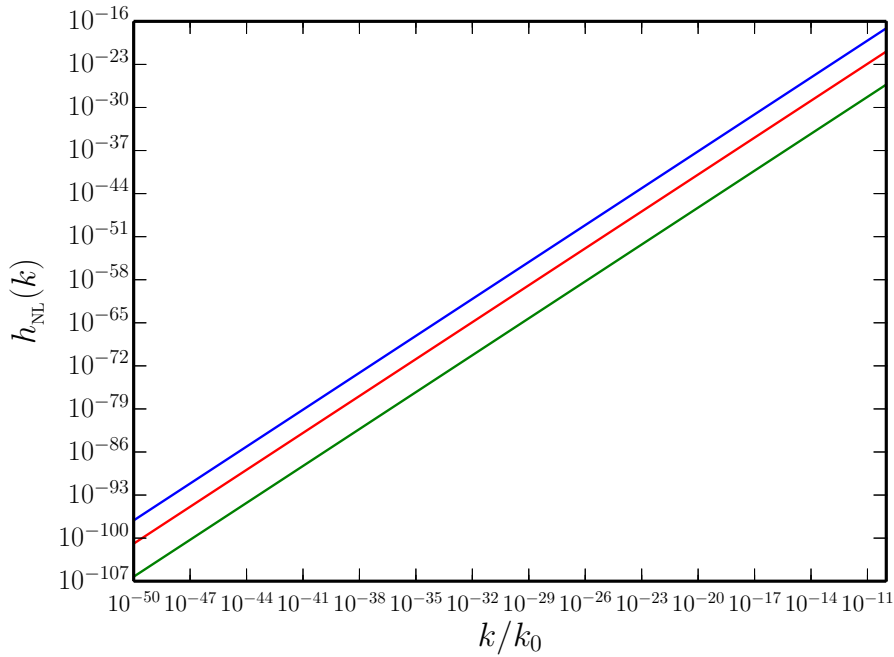}
\end{center}
\vskip -20pt
\caption{The contributions to the non-Gaussianity parameter $\hnl$ in the 
equilateral limit from the first (in green), the second (in red) and the 
third (in blue) domains have been plotted as a function of $k/k_0$ for a 
wide range of wavenumbers such that $k\ll k_0/\alpha$. 
We have worked with the same set of values as in the previous figure.
Clearly, the third domain gives rise to the maximum contribution to the
non-Gaussianity parameter $\hnl$.}
\label{fig:hnl-el}
\end{figure}

We can now make use of the behavior of the mode $h_k$ at $\ee=\beta\,\eta_0$ 
and substitute the results we have obtained above in the 
expressions~(\ref{eq:tbs}) and (\ref{eq:hnl}) to arrive at the tensor bi-spectrum 
and the corresponding non-Gaussianity parameter $\hnl$ for an arbitrary triangular 
configuration of the wavenumbers involved.
The resulting expressions prove to be rather long and, for this reason, we
shall illustrate the various results graphically for a set of suitable values 
of the parameters.
Let us first compare the contributions from the three domains.
Restricting ourselves to the equilateral limit, in Fig.~\ref{fig:hnl-el}, we 
have plotted the contributions to $\hnl$ from the three domains that we have 
considered. 
It is evident from the figure that the contribution due to the third domain 
to the parameter $\hnl$ turns out to be the maximum.
We find that the third domain contributes the maximum in the squeezed limit
as well.  

\par

\begin{figure}[!htbp] 
\begin{center}
\includegraphics[width=12.00cm]{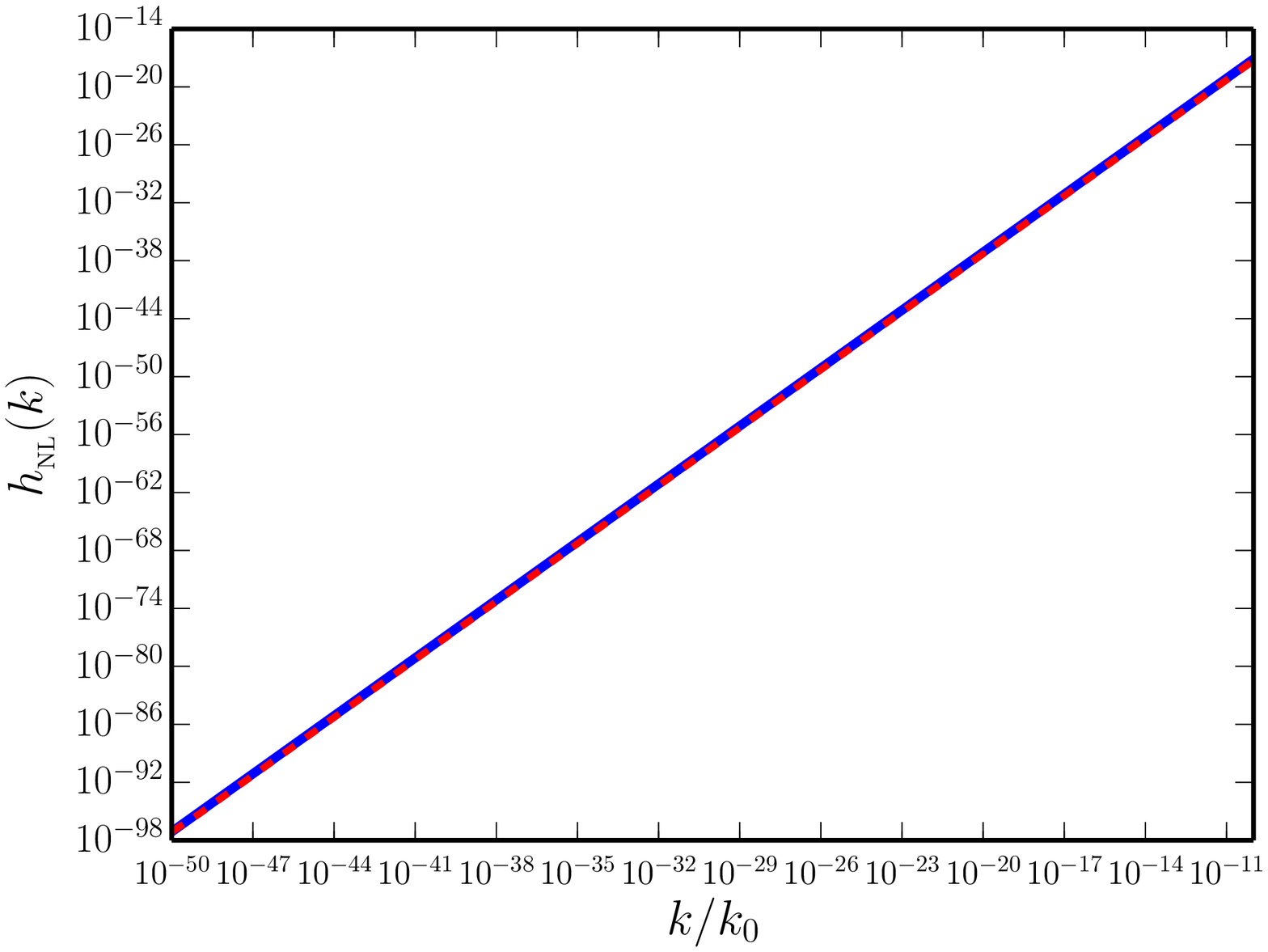}
\end{center}
\vskip -20pt
\caption{The behavior of the non-Gaussianity parameter $\hnl$ in the 
equilateral (in blue) and the squeezed (in red) limits have been 
plotted as a function of $k/k_0$ for a wide range of wavenumbers such
that $k\ll k_0/\alpha$. 
We have worked with the same set of values as in the earlier two figures.
Clearly, the resulting $\hnl$ is considerably small when compared to the 
values that arise in de Sitter inflation wherein $3/8\lesssim\hnl\lesssim 
1/2$.
Moreover, we find that $\hnl$ behaves as $k^2$ in the equilateral and the
squeezed limits, with similar amplitudes.
The fact that $\hnl$ is much smaller than $3/8$ in the squeezed limit
implies that the consistency condition is violated.}
\label{fig:hnl-el-sl}
\end{figure}

It is now a matter of adding the contributions from the three domains to arrive 
at the complete tensor bi-spectrum and the non-Gaussianity parameter $\hnl$.
In Fig.~\ref{fig:hnl-el-sl}, we have plotted the behavior of the 
non-Gaussianity parameter $\hnl$ in the equilateral and the squeezed 
limits. 
Three points concerning the figure require emphasis.
To begin with, we should mention that the non-Gaussianity parameter $\hnl$ 
behaves as $k^2$ in both the equilateral and the squeezed limits, with
virtually the same amplitude.
Secondly, the value of the parameter $\hnl$ is very small when compared to 
the values that occur in, say, de Sitter inflation wherein $3/8\lesssim\hnl
\lesssim 1/2$ (in this context, see Ref.~\cite{Sreenath:2013xra}).
Thirdly, since the tensor power spectrum is strictly scale invariant for
wavenumbers such that $k\ll k_0/\alpha$, the amplitude of the non-Gaussianity 
parameter $\hnl$ in the squeezed limit over such domain should be equal to 
$3/8$, if the consistency relation holds true (see Appendix, also see
Ref.~\cite{Sreenath:2014nka}).
Whereas, we find that $\hnl$ is considerably smaller than $3/8$ in the 
squeezed limit, which unambiguously implies that the consistency condition 
is violated.
Evidently, this behavior can be attributed to the fact that the amplitude
of the tensor mode does not freeze to a constant value at late times.  

\par

At this stage, we need to discuss the dependence of the tensor power and 
bi-spectra on the parameters $\alpha$ and $\beta$ that we have introduced.
We find that the tensor power spectrum and the bi-spectrum do not significantly
depend on $\alpha$ over a wide range of values, say, $10^5\lesssim \alpha \lesssim
10^{15}$.  
We had mentioned earlier that the tensor power spectrum has a rather weak 
dependence on $\beta$.
Whereas, we find that the tensor bi-spectrum grows roughly as $\beta^{3/2}$.
However, $\beta$ cannot be allowed to be too large for two reasons. 
One may a priori expect that the analytical approximation~(\ref{eq:hk-d2}) will 
remain valid until the time $-\eta_k=\sqrt{2}/k$ after the bounce.
We had pointed out that the evolution of the mode $h_k$ is asymmetric in $\eta$.
Actually, it can be shown that (using numerical analysis) the analytical 
approximation~(\ref{eq:hk-d2}) breaks down much before $-\eta_k$.
For this reason, we have chosen $\beta$ to be smaller than $\alpha$.
Moreover, a transition to the radiation dominated phase is expected to take place 
sometime after the bounce.
It seems reasonable to expect that such a transition will occur when the scale
factor is $a\simeq 10^4\, a_0$, which corresponds to $\beta=10^2$.
We find that our main conclusions, \viz that the value of $\hnl$ is small over  
cosmological scales and that the consistency relation is violated in the
squeezed limit, continue to remain valid even if we increase $\beta$ by, say,
a couple of orders of magnitude.

\par
 

\section{Discussion}\label{sec:d}

In this paper, we have analytically calculated the tensor bi-spectrum in a 
matter bounce using the Maldacena formalism.
While the matter bounce leads to a scale invariant tensor power spectrum for
scales of cosmological interest as de Sitter inflation does, we have shown 
that the non-Gaussianity parameter $\hnl$ that characterizes the amplitude
of the tensor bi-spectrum is much smaller than the corresponding values in
de Sitter inflation.
We have also shown that, due to the growth in amplitude of the tensor modes
as one approaches the bounce, the consistency condition is not satisfied by
the tensor bi-spectrum in the squeezed limit. 
Recall that, in the absence of detailed modelling of the bounce, we had 
assumed that $k_0 \simeq\Mpl$.
We should however clarify that, since $k\ll k_0$ for cosmological scales,
our essential conclusions, \viz that $\hnl$ is small and that the consistency 
condition is violated over such scales, will remain unaffected even if we 
choose $k_0$ to be a few orders of magnitude below the Planck scale.  
It will clearly be worthwhile to investigate these issues using a numerical 
approach in a wider class of bouncing models.

\par

In the bouncing scenario that we have considered, at very early times, 
\ie during the first domain of our interest, the contribution to the 
non-Gaussianity parameter $\hnl$ can be said to be small because the 
amplitude of the tensor perturbations themselves are small. 
In the second domain, although the scale factor decreases gradually to 
reach its minimum at the bounce, the non-Gaussianities become larger as 
the perturbations grow. 
In the third domain, \ie after the bounce, the scale factor increases 
steadily. 
Also, the amplitude of the perturbations do not freeze but grow slowly. 
Due to these reasons the contribution to the non-Gaussianity parameter 
is the largest from this regime.
However, essentially due to the form of the scale factor, one finds that
the parameter $\hnl$ has an overall $k^2$ dependence.
Since the scales of cosmological interest are about $50$ to $60$ orders 
below the Planck scale, the non-Gaussianity parameter $\hnl$ proves to
be very small over such scales.

\par

We believe that the results we have obtained have tremendous implications 
for the other three-point functions and, importantly, the scalar bi-spectrum.  
It seems clear that, due to the growth during the contracting phase near the
bounce, the consistency conditions governing the other three-point functions
will be violated as well~\cite{Cai:2009fn}. 
This possibly can act as a powerful discriminator between the inflationary
and bouncing scenarios.    
Within inflation, one requires peculiar situations to violate the consistency 
conditions~\cite{Namjoo:2012aa,Chen:2013aj}.
In contrast, in a bouncing scenario, the consistency relations seem to be
violated rather naturally.
Notably, situations involving violations of the consistency conditions have 
been considered as possible sources of spherical asymmetry in the early 
universe~\cite{Jeong:2012df,Dai:2013ikl,Dai:2013kra}.
These aspects seem worth exploring in greater detail.      


\section*{Acknowledgements}

The authors wish to thank Robert Brandenberger, Dhiraj Hazra, J\'er\^ome 
Martin and Patrick Peter for discussions and comments on the manuscript. 
VS would like to acknowledge Department of Science and Technology, India, 
for supporting this work through the project PHY1213274DSTXDAWO.
LS wishes to thank the Indian Institute of Technology Madras, Chennai, 
India, for support through the New Faculty Seed Grant.


\appendix

\section*{Appendix: The squeezed limit and the consistency condition}

An important property of the three-point functions is their behavior in the 
so-called squeezed limit~\cite{Maldacena:2002vr,Creminelli:2004yq,Cheung:2007sv,
RenauxPetel:2010ty,Ganc:2010ff,Creminelli:2011rh,Martin:2012pe,Sreenath:2014nca,
Kundu:2014gxa,Kundu:2015xta}.
As we have discussed before, the squeezed limit corresponds to the situation 
wherein one of the three wavenumbers is much smaller than the other two.
In such a limit, under certain conditions, it is known that all the three-point 
functions involving the scalars and tensors generated during inflation can be 
expressed entirely in terms of the two-point functions~\cite{Jeong:2012df,
Dai:2013ikl,Dai:2013kra,Kundu:2013gha,Sreenath:2014nka}.
In the context of inflation, these consistency conditions arise essentially
because of the fact that the amplitude of the long wavelength scalar and 
tensor modes freeze on super-Hubble scales.  
In this appendix, we shall outline the proof of the consistency condition
satisfied by the tensor bi-spectrum during inflation.

\par

Since the amplitude of a long wavelength mode freezes on super-Hubble scales 
during inflation, such modes can be treated as a background as far as the 
smaller wavelength modes are concerned.
Let us denote the constant amplitude of the long wavelength 
tensor mode as~$\gB_{ij}$.
In the presence of such a long wavelength mode, the background FLRW metric can 
be written as
\begin{equation}
\d s^2 = -\d t^2 + a^2(t)\, [{\rm e}^{\gBe}]_{ij}\,\d {\bm x}^i\,\d {\bm x}^j,
\end{equation}
\ie the spatial coordinates are modified according to a spatial transformation 
of the form ${\bm x'} = \Lambda\,{\bm x}$, where the components of the matrix 
$\Lambda$ are given by $\Lambda_{ij}=[{\rm e}^{\gBe/2}]_{ij}$.
Under such a spatial transformation, the small wavelength tensor perturbation 
transforms as~\cite{Sreenath:2014nka}
\begin{equation}
\g^\vk_{ij} \to {\rm det}\,(\Lambda^{-1})\,\g^{\Lambda^{-1}\,\vk}_{ij},
\end{equation}
where ${\det}\,(\Lambda^{-1})= 1$. 
Under these conditions, we also obtain that 
\begin{equation}
\vert\Lambda^{-1}\,\vk\vert=[1-\gB_{ij}\,k_i\,k_j/(2\,k^2)]\,k,
\end{equation}
where $k_i$ is the component of the wavevector ${\bm k}$ along the $i$-spatial 
direction and we have restricted ourselves to the leading order in $\gB_{ij}$.
Moreover, one can show that
\begin{equation}
\delta^{(3)}(\Lambda^{-1}\,{\bm k}_1+\Lambda^{-1}\,{\bm k}_2)
={\rm det}\,(\Lambda)\,\delta^{(3)}({\bm k}_1+{\bm k}_2)
=\delta^{(3)}({\bm k}_1+{\bm k}_2),
\end{equation}
since ${\det}\,(\Lambda)= 1$.
Upon using the above results, we find that the tensor two-point function in 
the presence of a long wavelength mode can be written as 
\begin{eqnarray}
\langle \hat{\g}^{\vka}_{m_1n_1}\,\hat{\g}^{\vkb}_{m_2n_2}\rangle_{k} 
&=& \f{(2\,\pi)^2}{2\,k_1^3}\,
\frac{\Pi^{\vka}_{m_1n_1,m_2n_2}}{4}\,\pt(k_1)\,
\delta^{(3)}(\vka + \vkb)\nn\\
& &\times\,\l[ 1 - \l(\frac{\nt-3}{2}\r)\,
\gB_{ij}\,{\hat n}_{1i}\,{\hat n}_{1j}\r].
\end{eqnarray}
where ${\hat n}_{1i}=k_{1i}/k_1$ and the long wavelength mode is denoted by 
the wavenumber $k$, while $\nt$ represents the tensor spectral index. 
The corresponding expression for the tensor bi-spectrum can be obtained from
the above result to be
\begin{eqnarray}
\langle\,\hat{\g}^{\vka}_{m_1n_1}\,\hat{\g}^{\vkb}_{m_2n_2}\, 
\hat{\g}^{\vkc}_{m_3n_3}\,\rangle_{k_3} 
&\equiv&\langle\,\langle\,\hat{\g}^{\vka}_{m_1n_1}\, 
\hat{\g}^{\vkb}_{m_2n_2}\,\rangle_{k_3}\,
\hat{\g}^{\vkc}_{m_3n_3}\,\rangle \nn\\
&=&-\,\f{(2\,\pi)^{5/2}}{4\,k_1^3\, k_3^3}\,\l(\f{\nt-3}{32}\r)\,
\pt(k_1)\,\pt(k_3)\nn\\
& &\times\,\Pi^{\vka}_{m_1n_1,m_2n_2}\,\Pi^{\vkc}_{m_3n_3,ij}\,
{\hat n}_{1i}\,{\hat n}_{1j}\,
\delta^{3}(\vka + \vkb),
\end{eqnarray}
where ${\bm \vkc}$ has been considered to be the squeezed mode.
The above relation wherein the tensor bi-spectrum has been expressed 
completely in terms of the power spectrum is known as the consistency 
condition~\cite{Kundu:2013gha,Sreenath:2014nka}.
Upon substituting this expression in the definition for the tensor 
non-Gaussianity parameter $\hnl$ [cf.~Eq.~(\ref{eq:hnl-ov})], we 
find that we can express the consistency relation in the squeezed 
limit as follows:
\begin{eqnarray}
 \lim_{k_3\to 0}\,\hnl(\vk,-\vk,\vkc) 
&=& \l[\f{\nt(k) - 3}{2}\r]\, 
\biggl(2\,\Pi^{\vk}_{m_1n_1,m_2n_2}\,\Pi^{\vkc}_{m_3n_3,{\bar m}{\bar n}} 
+ \Pi^{\vk}_{m_1n_1,{\bar m}{\bar n}}\,\Pi^{\vkc}_{m_3n_3,m_2n_2}\nn\\ 
& &+\,\Pi^{\vk}_{{\bar m}{\bar n},m_2n_2}\,\Pi^{\vkc}_{m_3n_3,m_1n_1}\biggr)^{-1}\,
\Pi^{\vk}_{m_1n_1,m_2n_2}\, \Pi^{\vkc}_{m_3n_3,ij}\,{\hat n}_{i}\,{\hat n}_{j}.
\end{eqnarray}
Actually, an overall minus sign occurs in the expression for $\hnl$ in the
squeezed limit due to the polarization tensors~\cite{Sreenath:2014nka}, 
Therefore, if we ignore the polarization tensors, in the domain where the 
tensor power spectrum is strictly scale invariant (\ie when $\nt=0$), the 
value of $\hnl$ in the squeezed limit reduces to~$3/8$, if the consistency 
relation is satisfied.


\bibliographystyle{JHEP}
\bibliography{tbs-mb-september-2015}


\end{document}